# Mitigating Position Bias with Regularization for Recommender Systems

Hao WANG[1]
*CEO & Founder, Ratidar Technologies LLC, China*

**Abstract.** Fairness is a popular research topic in recent years. A research topic closely related to fairness is bias and debiasing. Among different types of bias problems, position bias is one of the most widely encountered symptoms. Position bias means that recommended items on top of the recommendation list has a higher likelihood to be clicked than items on bottom of the same list. To mitigate this problem, we propose to use regularization technique to reduce the bias effect. In the experiment section, we prove that our method is superior to other modern algorithms.

## 1. Introduction

Recommender systems have been evolving for a couple of decades. The technology has been widely applied in commercial products such as TikTok, Amazon and Kuai Shou. The major benefit of the technology is that by hiring a small team of experts and purchasing the necessary IT infrastructures, the company will be able to attract a large number of visitors or sales volume, and saving millions or even billions of money on marketing. The increase of user traffic or sales volume could go up to as high as 30% to 40%.

Recommender system has some intrinsic problems such as bias problems. The following bias problems have been identified in recommender systems in recent years: Popularity Bias, Selection Bias, and Position Bias, etc. Popularity Bias refers to the effect that popular items are more likely to be recommended than less popular items in a disproportional way that could deeply effect the recommendation results. For example, when we apply collaborative filtering algorithms to solve book recommendation problem, popular books such as The Three Body Problem is read by many readers, and can bias the recommendation results due to the similarity computation procedures. Selection Bias refers to the effect that users tend to rate items that are extremely favored or extremely disliked by themselves. Position Bias happens when users prefer to click items recommended on top of the list while overlooking items listed at the bottom.

To mitigate the bias problems, a series of new approaches have been invented. Among different algorithms, popularity bias is the most well studied topic. Researchers in Google came up with a regularization technique named Focused Learning [1] while other researchers invented algorithms such as Zipf Matrix Factorization [2] and KL-Mat [3]. Since regularization is such a successful technique, in this paper, we apply the technique to mitigate the position bias of recommender system.

---

[1] Corresponding author: haow85@live.com

## 2. Related Work

Recommender system is one of the most common technologies in internet companies. For large-scale internet corporations, recommender system can greatly boost traffic and sales volume. Recommender systems have evolved from early shallow models such as collaborative filtering [4] and matrix factorization [5][6] to modern day models such as DLRM [7].

Algorithmic fairness (by fairness, we also mean bias and debiasing) is a well-known problem intrinsic to recommender system since the beginning. In 2018, Google proposed a fair recommender system called Focused Learning [1] by penalizing the loss function of matrix factorization with a penalty term. Two years later, a new fair algorithm named MatRec [8] was invented to tackle the fairness problem by incorporating popularity ranks into the loss function formulation. In 2021, Wang introduced an algorithm named Zipf Matrix Factorization [2] that uses a penalty term named Degree of Matthew Effect [2] to penalize the loss function of matrix factorization. KL-Mat was proposed in the following year as a different fair algorithm that also relies on the framework of matrix factorization [3].

In this paper, we also use the framework of matrix factorization to develop our new algorithm which mitigates the position bias effect. Matrix factorization has many variants such as SVDFeature [9], Alternating Least Squares [10], SVD++ [11], timeSVD [12] and so on. Due to its simplicity and interpretability, matrix factorization is a common used technical approach in the industry.

## 3. Matrix Factorization

Matrix Factorization defines the recommendation problem as the following problem:

$$L = \sum_{i=1}^{n} \sum_{j=1}^{m} \left( R_{i,j} - U_i^T \cdot V_j \right)^2$$

The problem formulation could be viewed as an angle- preserving dimension reduction problem, as explained in [13], if we revise the formulation a bit (as the common practice in the real world):

$$L = \sum_{i=1}^{n} \sum_{j=1}^{m} \left( \frac{R_{i,j}}{R_{max}} - \frac{U_i^T \cdot V_j}{||U_i^T \cdot V_j||} \right)^2$$

Matrix Factorization could be solved for its optimal parameters using optimization techniques such as Stochastic Gradient Descent (SGD).

To solve the bias problems such as popularity bias, many research works have used matrix factorization as the test-bed for debiasing problems. We briefly review Zipf Matrix Factorization and KL-Mat in this section since their derivation is similar to our newly proposed algorithm.

Zipf Matrix Factorization was introduced in 2021. The basic idea is to use a penalty term named Degree of Matthew Effect to penalize the classic matrix factorization loss function:

$$L = \sum_{i=1}^{n}\sum_{j=1}^{m}\left(\frac{R_{i,j}}{R_{max}} - \frac{U_i^T \cdot V_j}{||U_i^T \cdot V_j||}\right)^2 - \beta\left(1 + n\left(\sum_{i=1}^{n} \ln \frac{x_i}{x_{max}}\right)^{-1}\right)$$

The regularization term is named Degree of Matthew Effect by researchers. The x's in the formula represent the ranks of items in the recommendation result list. To estimate the ranks, two rounds of regressions are applied to first estimate the linear coefficients of U and V in the equation systems where ranks are approximated by linear combinations of U and V dot products. In the second round, coefficients computed in the first round are set to constants while new values of U and V are computed.

KL-Mat is an invention first published in 2022. The idea is to use an important concept named KL-Divergence in the field of Information Geometry. The loss function for KL-Mat is formulated as follows:

$$L = \sum_{i=1}^{n}\sum_{j=1}^{m}\left(\frac{R_{i,j}}{R_{max}} - \frac{U_i^T \cdot V_j}{||U_i^T \cdot V_j||}\right)^2 + \beta\left(\sum_{j=1}^{n} p\left(\frac{1}{rank_j}\right)\log\left(\frac{p(rank_j)}{q(rank_j)}\right) + \sum_{j=1}^{n} q\left(\frac{1}{rank_j}\right)\log\left(\frac{q(rank_j)}{p(rank_j)}\right)\right)$$

The penalty term is KL-Divergence that computes the difference between the real item popularity rank distribution and the uniform distribution. Just like Zipf Matrix Factorization, researchers use linear combination of dot products of U and V to approximate the ranks. We apply Stochastic Gradient Descent (SGD) to solve the problem and the experimental results are competitive with other modern day approaches.

**4. Position Bias Solution**

It is well known that recommender system has position bias problem. Position bias means that when items are recommended to the user, items on the top of the list is more likely to be clicked than items at the bottom. A better recommender system would have less position bias issue while at the same time maintaining the high accuracy of the recommender system results.

To mitigate the position bias problem, we propose to use regularization techniques with a newly designed penalty term. The loss function is defined below:

$$L = \sum_{i=1}^{n}\sum_{j=1}^{m}\left(\frac{R_{i,j}}{R_{max}} - \frac{U_i^T \cdot V_j}{||U_i^T \cdot V_j||}\right)^2 + \beta\left(\sum_{i=1}^{n}\sum_{j=1}^{m}\left(\frac{1}{\text{position}_{i,j}} - \frac{1}{m}\right)\right)^2$$

The intuition behind this Formula is that the position bias mostly follows power law distribution. To simplify our assumption even further, we assume the position clicks follow Zipf Distribution. Namely the number of clicks on the first item is T, and the number of clicks on the second item is T/2, ... Since the position on the list is ranked by the user item rating score, namely R, we could reformulate the loss function as below:

$$L = \sum_{i=1}^{n}\sum_{j=1}^{m}\left(\frac{R_{i,j}}{R_{max}} - \frac{U_i^T \cdot V_j}{||U_i^T \cdot V_j||}\right)^2 + \beta\left(\sum_{i=1}^{n}\sum_{j=1}^{m}\left(\frac{R_{i,j}}{R_{max}} - \frac{1}{m}\right)\right)^2$$

The penalty term is built upon the following theory: For a recommender system without position bias problem, click numbers of the items on the recommendation list should follow uniform distribution. In other words, the probability of being clicked follows $\frac{1}{m}$, where m is the number of items in total. If we take a closer look at the penalty term, when user item rating is large, $\frac{R_{i,j}}{R_{max}}$ approaches 1, and therefore the penalty term has a larger value than when user item rating is small. The task of the penalty term, therefore, is to penalize position bias caused by large user item rating values more than bias caused by small user item rating values. Since large user item rating values are usually associated with large number of clicks or likes, the penalty term is biased to penalize position bias of the majority of clicks.

In order to solve the loss function for the optimal parameter values of U and V, we resort to Stochastic Gradient Descent (SGD) and obtain the following parameter update rules:

$$\frac{\partial L}{\partial U_i} = \frac{2\beta(t_3 - t_6)}{t_2}V_j - \frac{2\left(\frac{R_{i,j}}{R_{max}} - t_3\right)}{t_2}V_j + \frac{2t_4\left(\frac{R_{i,j}}{R_{max}} - t_5\right)}{t_7}U_i - \frac{2\beta t_4(t_5 - t_6)}{t_7}U_i$$

where:

$$t_0 = ||U_i||_2, \quad t_1 = ||V_j||_2, t_2 = t_0 \cdot t_1$$

$$t_3 = \frac{U_i^T \cdot V_j}{t_2}$$

$$t_4 = V_j^T \cdot U_i, t_5 = \frac{t_4}{t_2}$$

$$t_6 = \frac{1}{m}, t_7 = t_1 \cdot t_0^3$$

The partial derivative of the loss function with respect to V is computed as below:

$$\frac{\partial L}{\partial V_j} = \frac{2\beta t_6}{t_2}U_i - \frac{t_5}{t_2}U_i + \frac{t_3 \cdot t_5}{t_7}V_j - \frac{2\beta t_3 \cdot t_6}{t_7}V_j$$

The pseudo-code for the algorithm is illustrated below:

$$t_0 = ||U_i||_2, t_1 = ||V_j||_2$$

$$t_2 = t_0 \cdot t_1, t_3 = V_j^T \cdot U_i$$

$$t_4 = \frac{t_3}{t_2}, \quad t_5 = 2\left(\frac{R_{i,j}}{R_{max}} - t_4\right), \quad t_6 = t_4 - \frac{1}{m}, \quad t_7 = t_0 \cdot t_1^3$$

## 5. Experiments

We test our algorithm on 2 different data sets: MovieLens 1 Million Dataset [14] and LDOS-CoMoDa Dataset [15]. The first data set contains 6040 users and 3706 items; the second data set includes 121 users and 1232 items.

The algorithms in our experimental comparison are ZeroMat [16], Random Placement, Classic Matrix Factorization, DotMat [17], DotMat Hybrid [17], Zipf Placement, PoissonMat [18], PoissonMat Hybrid [18] and Position Bias (Our new algorithm).

ZeroMat, DotMat, PoissonMat are zeroshot-learning algorithms aiming to solve the cold-start problem for recommender systems. DotMat Hybrid and PoissonMat Hybrid are hybrid models that build upon DotMat and PoissonMat, both of which yield very competitive results, as explained in [17] and [18]. Random Placement and Zipf Placement are basically random recommendation heuristics using uniform or Zipf distribution.

The evaluation metrics we use are MAE score, Degree of Matthew Effect and Position Bias Metrics. MAE score is a commonly used accuracy metric, so we do not elaborate here. Degree of Matthew Effect is a new fairness metric introduced in [2]. Position Bias Metrics is essentially the penalty term of our new algorithm divided by the total number of user item rating values in the test dataset.

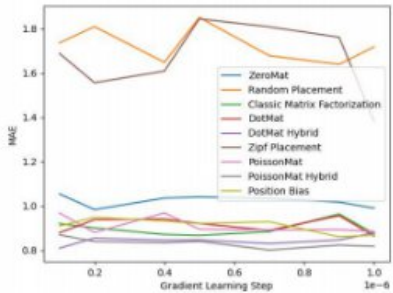

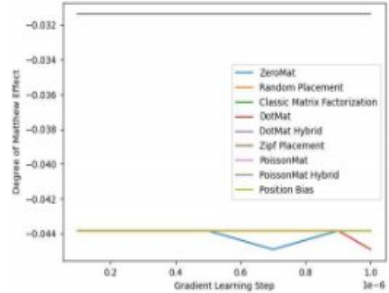

Fig. 1 Accuracy Comparison on MovieLens 1 Million Dataset

Fig. 2 Popularity bias Comparison on MovieLens 1 Million Dataset

From Fig. 1 to Fig. 3 we observe that for MovieLens 1 Million Data Set, our new algorithm (named Position Bias in the figures) is not the best algorithm on MAE score and Degree of Matthew Effect metric, but is highly competitive when evaluated by metric for Position Bias.

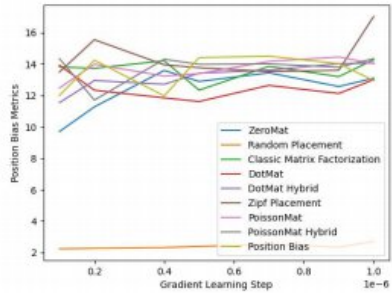

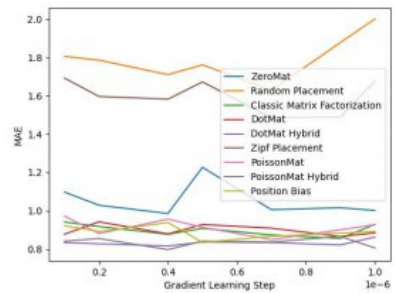

Fig. 3 Position Bias Comparison on MovieLens 1 Million Dataset

Fig. 4 MAE Comparison on LDOS-CoMoDa Dataset

Fig. 4 to Fig. 6 illustrate the algorithm performance comparison on LDOS-CoMoDa data set. We notice that our new algorithm is competitive with other algorithms on MAE and Degree of Matthew Effect scores at its optimum, and also it is one of the best algorithm when tested on Position Bias Metric Score. We also observe a very interesting phenomenon, that is ZeroMat is the best algorithm when evaluated on Position Bias Metric Score.

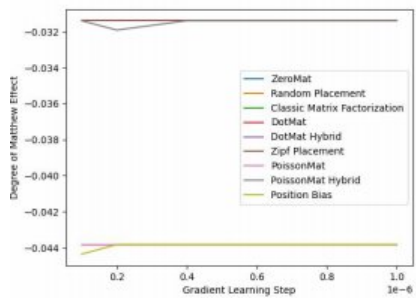
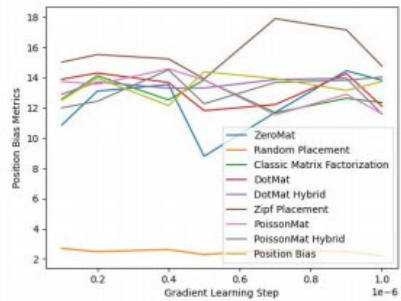

Fig. 5 Popularity Bias Comparison on LDOS-CoMoDa Dataset

Fig. 6 Position Bias Comparison on LDOS-CoMoDa Dataset

## 6. Conclusion

In this paper, we proposed a new algorithm that solves the position bias problem in recommender systems. We resort to regularization technique with a carefully designed penalty term with simplicity and interpretability. In the experiment section, we prove that our algorithm is competitive with modern day recommender system algorithms. In future work, we would like to explore solutions to other bias problems such as selection bias and exposure bias. We would like to enhance the fairness of AI algorithms in our future works.